\documentclass[letterpaper, 12pt, twocolumn]{article}

\usepackage[T1]{fontenc}

\usepackage{geometry}
\usepackage{setspace}

\usepackage[style = chem-acs, articletitle=true]{biblatex}
\usepackage{graphicx}
\usepackage{float}

\newfloat{scheme}{htbp}{los}
\floatname{scheme}{Scheme}
\floatname{chart}{Chart}
\newfloat{graph}{htbp}{loh}

\usepackage{chemformula} 
\usepackage{comment}
\usepackage{graphicx}
\usepackage{xcolor}
\newcommand{\new}[1]{\textcolor{black}{#1}}
\newcommand{\newer}[1]{\textcolor{black}{#1}}
\newcommand{\EA}{E_A}
\usepackage{authblk}
\author[1]{Vanessa K. Ward*} 
\author[1]{Karen E. Johnston}
\affil[1]{Department of Chemistry, Durham University, South Road, Durham, DH1 3LE, United Kingdom}

\title{A Rule That's Made to be Broken? Reframing the \mbox{Arrhenius} Law for Ionic Transport}
\date{*Email: vanessa.k.ward@durham.ac.uk}

\begin{document}
\maketitle

\begin{abstract}
The Arrhenius law is ubiquitous in the physical sciences. It describes how an observable property scales with temperature, $T$, as $\exp(\pm\EA/k_BT)$ for a constant $\EA$, the `activation energy'. Fitting to the Arrhenius law is a common procedure for both experimental and computational data. However, the physical significance of any `activation energy' calculated is unclear. Here we focus on self-diffusion coefficients for potential solid electrolytes, and by identifying the individual `diffusive steps' within the system, show how a physically significant activation energy can be extracted. There are also many examples of non-Arrhenius behaviour including `super-Arrhenius' curvature, where the effective activation energy appears to increase with decreasing temperature. We demonstrate how it is still possible to extract the underlying Arrhenius law, by quantifying self-correlations such as back-and-forth motion. The methodology described here is material-agnostic and can be utilised to analyse the molecular dynamics trajectories of both ordered and disordered materials.
\end{abstract}





\begin{center}
\begin{minipage}{3.25in}
  \sffamily
  \frenchspacing
  \includegraphics[]{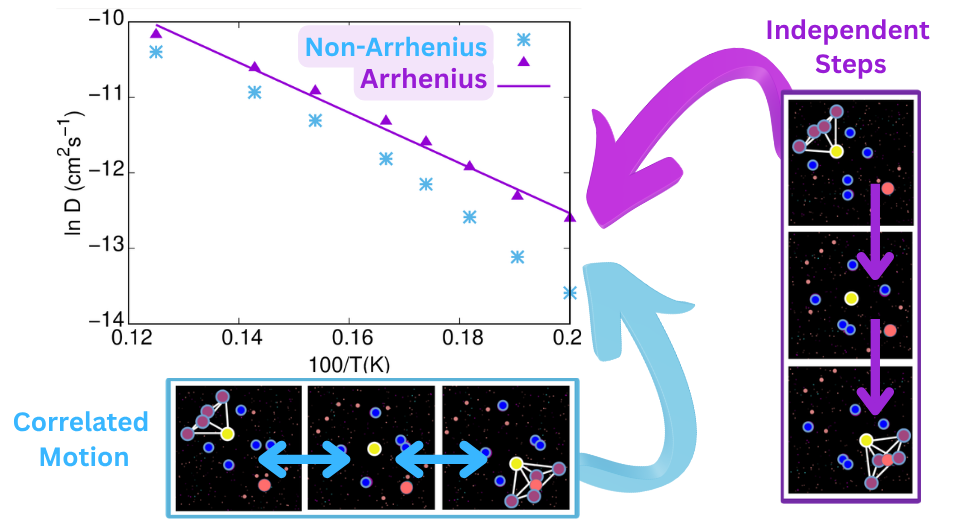}
\end{minipage}
\end{center}


The systematic study of the kinetics of chemical reactions in the 1800s revealed an appreciable dependence on temperature, leading to the familiar Arrhenius law, $k(T)=A\exp(-\EA/k_BT)$, where $k_B$ is the Boltzmann constant, and $A$ and $\EA$ are constants for the particular reaction. $\EA$ was labelled the `Activation Energy' on the basis that it represented the energy required to convert the reacting substance into the `active' form~\autocite{Logan1982Arrhenius}.

The Arrhenius law is used routinely in both the physical sciences and biology~\autocite{kohout2021modified}. It is regularly applied to fit both experimental and simulation data, including coefficients for transport processes such as viscosity, conductivity and diffusivity. Over a limited range of temperature, the logarithm of a transport coefficient plotted against $1/T$ is generally close to linear, indicative of an Arrhenius temperature dependence. 

However, over wider temperature ranges, data on Arrhenius plots is often found to be curved~\autocite{Logan1982Arrhenius,noori2024activation}. One interpretation of this curvature is that $\EA$ changes with temperature.  In particular, super-Arrhenius behaviour~\autocite{angell1991StrongFragile}, a convex curve on an Arrhenius plot, has been identified in numerous amorphous materials as well as molten salts and concentrated aqueous solutions. In terms of the Arrhenius equation, the interpretation is that the activation energy increases as temperature falls. Modified equations have also been developed with a constant activation energy but a divergence before $T=0$~\autocite{angell2000relaxation, kohout2021modified}.  Another alternative fitting procedure, is to fit multiple quasi-Arrhenius forms to the data. Multiple fits would indicate a change of mechanism with distinct activation energies for each fitted region~\autocite{kohout2021modified, qi2021bridging}. 

As fitting to Arrhenius law is a standard procedure, a better understanding of the origin of $\EA$ would be very valuable. In this work, we focus on self-diffusion coefficients, or the tracer diffusion if measured experimentally. Self-diffusivity is related to conductivity via the Nernst-Einstein relation and the Haven ratio~\autocite{marcolongo2017ionic,vargas-barbosar2020}. Although the methodology used here is material-agnostic, it has particular relevance for optimising ionic conductivity in solid electrolytes. Due to the short-time limitations of \textit{ab-initio} molecular dynamics, simulations are often run at high temperature, relying on an Arrhenius law extrapolation for room temperature predictions~\autocite{gao2020classical}. These extrapolations may have limited validity~\autocite{mccluskey2026bayesian}. For some ionic conducting materials, it has even been suggested that a non-Arrhenius temperature dependence is essential for optimising conductivity~\autocite{Kincs1996FastIonic}.  

\begin{figure}[h]
\centering
  \includegraphics[height=3.5cm]{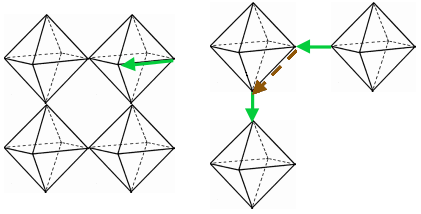}
  \caption{Schematic structures with ion sites shown by the vertices of each octahedron. The arrows show possible `hops' for the ions, moving between sites. On the left is a crystalline structure (Li$_3$OCl), with all vertices connected to other octahedra. On the right is a more flexible structure (Li$_6$OCl$_4$), where the octahedra are not connected.}
  \label{fig1}
\end{figure}

Particle motion in solids can often be described by distinct, discrete events or `hops'~\autocite{morgan2017}. The simplest models of diffusion involve random walkers jumping between sites. A random walk diffusion coefficient can be calculated as $D_{rw}=\alpha^2\nu/6$, where $\alpha$ is the hopping distance and $\nu$ is the hop frequency. Figure~\ref{fig1} shows some examples of how hopping could move particles from site to site in schematic structures. Certain conditions validate the use of a hopping model~\autocite{catlow1983static}. Firstly, the time between events ($1/\nu$) is much longer than the event duration ($t_{event}$), $1/\nu \gg t_{event}$. Secondly, the event distance ($\alpha$) is considerably longer than the amplitude of thermal motions ($a$), $\alpha\gg a$. Lastly, the maximum change in potential energy ($\Delta E$) is significantly larger than $k_BT$. As the temperature increases, the system may have enough energy to move freely over the barriers and a hopping model is no longer relevant. In the same way, the hopping model may also break down for some `superionic' electrolytes that exhibit very high conductivity, where ions follow a more liquid-like diffusion~\autocite{morgan2017}. 

In solids, particles are surrounded by nearest-neighbour `cages’~\autocite{siegel2021establishing}.  If the particle cannot escape from this cage, there is no contribution to net diffusion~\autocite{desouza2008energy}. For crystalline materials, cages are formed by the lattice structure and determining site occupancy is relatively easy~\autocite{deKlerk2018analysis}. Cage escapes can then be identified as an ion moving to a neighbouring site, see Figure~\ref{fig1}. In disordered materials, more sophisticated methods are needed. Firstly, a particle's cage in a disordered material must be defined, for example by voronoi analysis~\autocite{okada2015non}, solid-angle based algorithms~\autocite{van2012parameter}, or a nearest-neighbour analysis~\autocite{desouza2008energy, niblett2016dynamics, niblett2017pathways}. The second challenge is to identify {\it cage escapes}, differentiating them from smaller amplitude vibrations~\autocite{vollmayr2004single, kluge2004diffusion}. In disordered materials there are wide distributions of movements, including those that are less clearly identified as either vibrations or ‘hopping’. In these cases, analysing changes in nearest neighbours can be used as a signature for cage escapes~\autocite{desouza2008energy}. Even for crystalline materials, it is useful to have a methodology that does not depend on a fixed lattice structure and prior knowledge of the ion hops. The main transport mechanisms in crystalline materials involve point defects~\autocite{fu2023review} that can distort the lattice structure, and pathways typically include intermediate sites~\autocite{lu2020superionic}. 

Transport is a collective property, and unless the system is very dilute, involves dynamic correlations between ion movements~\autocite{vargas-barbosar2020}. For diffusion, the relevant correlations are self-correlations, where the movement of an ion depends on the direction of its previous movements. These correlation effects play an important role in solid-state diffusion~\autocite{philibert1991atom}. Back-and-forth jumps, for example, where an ion preferentially returns to its previous site, would cause a random walk diffusion coefficient ($D_{rw}$) to greatly overestimate diffusion, particularly at lower temperatures. Self-correlations will depend on both the solid structure and mechanisms for diffusion, and how these vary with temperature. They can be characterised by a correlation factor, $f$, such that $D=f\alpha^2\nu/6$, where $f=1$ corresponds to a random walk process and for $f=0$, there is no net diffusion.

In the case of vacancy diffusion in a lattice, the self-correlation factor is directly related to how the lattice geometry restricts the number of available pathways. It is possible to calculate the expected $f$ for vacancy diffusion in different crystal lattice geometries~\autocite{koiwa1983random, morgan2017,chen2017datamining}, and also for different diffusion mechanisms within those lattices~\autocite{gao2020classical}. The concentration of \new{vacancies} in a lattice structure will also affect the self-correlations. If the \new{vacancy} concentration increases, self-correlations decrease, and the deviation from random walk behaviour is reduced~\autocite{morgan2017}. Although harder to identify, back-and-forth motion can also be characterised in amorphous structures~\autocite{vollmayr2004single,desouza2008energy}, and a decrease in $f$ is expected if the number of available pathways decreases, for example at higher densities or lower temperatures. Examination of the simulation data for lithium superionic conductors showed that a marked decrease in the number and diversity of diffusion pathways at lower temperatures could be associated with super-Arrhenius behaviour~\autocite{qi2021bridging}. 


A common route for predicting the conductivity of a new material computationally is to calculate self-diffusion activation energies, $\EA$, in a two-fold manner. The first method is via a nudged-elastic-band (NEB) calculation, providing energy barrier details for a single hopping event using a detailed \textit{ab-initio} calculation on a small system~\autocite{gao2020classical, meier2014solid}. However, this method relies on an initial pathway guess, with fixed start and end points, and generally considers one event in isolation. The second method is to fit an Arrhenius law to longer-time averaged data from molecular dynamics (MD) simulations~\autocite{frenkel2023understanding, allen2017computer} (or experiments) over a range of temperatures. This method is limited by the validity of the Arrhenius law and the fitted $\EA$. The methodology described here provides a bridge between these two methods, analysing MD trajectories and extracting pathway information. The extracted information will be limited by the simulation potential but the method can be used to analyse any MD or \textit{ab-initio} MD simulations including those using machine-learned potentials. This work demonstrates the method by analysing the data presented in Reference~\parencite{dutra2023defect}, where LAMMPS MD simulations~\autocite{LAMMPS} were used to obtain self-diffusion coefficients of an anti-perovskite, Li$_3$OCl, and the related low-dimensional-networked Li$_6$OCl$_4$. \new{\textit{Ab-initio} simulations and NEB barrier calculations for smaller-scale systems are available~\autocite{lu2020superionic} and also previous MD simulations for Li$_3$OCl~\autocite{dawson2018screening}.} \new{Further simulation details are given in the Supplementary Information (SI).} Despite structural similarities, the diffusion coefficients for these two materials differ by several orders of magnitude. Their behaviour with temperature is also contrasting, Li$_3$OCl follows an Arrhenius law and Li$_6$OCl$_4$ displays super-Arrhenius curvature. Previous work showed that defects are required for the lithium ions to diffuse \new{in Li$_3$OCl} and that the dominant defects were lithium-chloride Schottky pairs~\autocite{dutra2023defect}. \new{For this reason, a $\sim 2\%$ concentration of lithium-chloride Schottky defects was imposed in the original simulations.} 

In this work, ion hopping and self-correlation events are identified within the MD trajectories for Li$_3$OCl and Li$_6$OCl$_4$. \new{To demonstrate applicability of the method across multiple systems, the related Li$_3$OBr and Li$_6$OBr$_4$ systems were also investigated and are shown in the SI.} To identify hopping events in both the crystalline Li$_3$OCl and flexible Li$_6$OCl$_4$ structures, a nearest-neighbour analysis is undertaken by first quenching each configuration to the potential energy surface. Within this potential energy landscape (PEL), any significant changes are much more readily identified as thermal noise has been removed. The first peak of the radial distribution function is very well-defined and can be used to identify the nearest neighbours for each lithium ion. A distance criterion can then be used to identify neighbours lost/gained due to movement away from this peak and hence isolate cage escape events. The results are insensitive to the precise value of this distance criterion~\autocite{desouza2008energy}. The angle, $\theta$, between successive hops is used to recognise any correlated events. $|\cos\,\theta|\geq 0.9$ identifies collinear events, but as with the previous distance criterion, the precise value of this cutoff makes little difference to the results. This PEL-enabled nearest-neighbour method (EL-Cage~\autocite{elcage}) for identifying hopping events and correlated motion is broadly applicable~\autocite{desouza2008energy, niblett2016dynamics, niblett2017pathways}. Expertise is required for predicting the pertinent neighbour changes for cage-breaking events in a particular material and for determining the frequency of the trajectory information required to successfully identify these events. However, there is a clear feedback loop for validation, as discrepancies between diffusion coefficients from EL-Cage and those obtained from MD can be used to refine and improve the method.

\begin{figure}[h]
\centering
  \includegraphics[height=5.2cm]{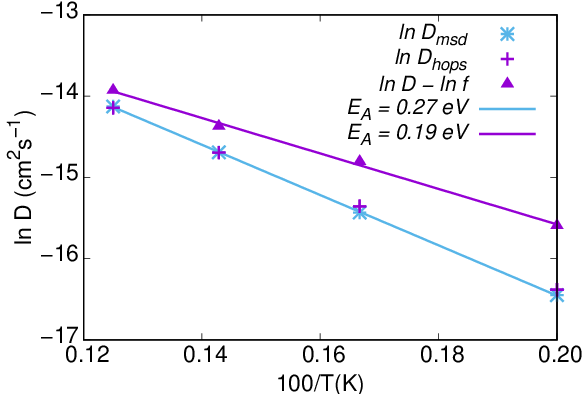}
  \caption{Diffusion coefficients for Li$_3$OCl. The blue line shows an Arrhenius fit to the original diffusion coefficients calculated from the trajectory mean-square displacement and the purple line shows a fit to the `random walk' diffusion coefficients ($D_{rw}$) with correlation effects removed.}
  \label{fig2}
\end{figure}
\begin{figure}[h]
\centering
  \includegraphics[height=2.3cm]{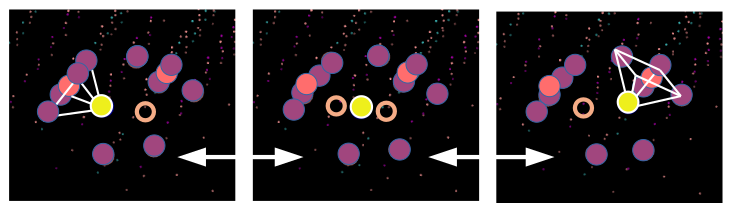}
  \caption{The detected movements of a lithium ion (yellow) from one lattice site (left) via an interstitial site (middle) to another lattice site (right). These potential-energy-minimised structures are taken from an MD trajectory. Vacant sites (orange circles), lithium (purple) and oxygen (pink) neighbouring ions are shown, with chlorines omitted for clarity.}
  \label{fig3}
\end{figure}

The Li$_3$OCl anti-perovskite structure consists of six lithium ions surrounding an oxygen in an octahedral arrangement, Figure~\ref{fig1} (left). These octahedra are connected by their vertices in all three dimensions such that each lithium ion has two oxygen neighbours. Chlorines fill the spaces between the octahedra. For this ordered crystal structure, diffusion coefficients calculated from the mean-square displacement appear to follow Arrhenius law, see Figure~\ref{fig2} (blue, lower line).  \new{An equivalent plot for the related Li$_3$OBr system is also shown in the SI, Figure~S2.} To isolate lithium-ion cage escapes, changes in their oxygen neighbours are detected. \newer{Although free to move, the oxygens generally remain} in their lattice positions throughout the trajectory. The expected hopping distance from one lithium site to another, would be $\sim 2.8\,\text{\r{A}}$ \autocite{dutra2023defect}. Although hops of this distance are detected, many events only correspond to less than half this distance, indicating an interstitial site. Figure~\ref{fig3} shows an example of the vacancy-interstitial-vacancy mechanism from a PE-minimised simulation trajectory.

Movements from a lattice site to an interstitial position can be detected by the loss of a single oxygen neighbour. The gain of a different oxygen neighbour then identifies the continuation hop from an interstitial position to a lattice site. For ease of analysis, if any direct site-to-site hops are detected, they are decomposed into two steps. \new{The analysis can also be undertaken for direct site-to-site hops rather than the two step process, as shown in the SI, Figure~S3.}  In practice, the lithium ion is moving to a vacant site and the presence of the vacancy means that the rest of the ions are slightly displaced from their original positions, attempting to compensate for the missing ion. The total distance to move from one site to another, $\sim2.4\,\text{\r{A}}$, is therefore slightly smaller than the average site-to-site distance. 

The correlation factor, $f$, is determined by counting chains of correlated events, both direct reversals and continuation hops, see Equation~\ref{eq:fh3}. \new{Each hop, $i$, is assigned a directional number, $d_i$,} with a value of $\pm1$ depending on how it is correlated to the first hop, creating a one-dimensional description of each chain of events. All hops are assumed to be equal in length and it is assumed that the only correlations between hops are movements between collinear points within a chain. \begin{equation}\label{eq:fh3}
\begin{aligned}
f&=\frac{1}{N}\sum_M\sum_{i,j}d_{i}^{(M)}d_{j}^{(M)},
\end{aligned}
\end{equation}
where $i,j$ are indices for the hops within a chain of events. $M$ is the chain index and $N$ is the total number of events. Directional numbers, $d_i$, for events within different chains, including any uncorrelated hops, are orthogonal. The only non-zero cross-terms are between hops within a particular chain. The values of $f$ for the full trajectory vary from $0.45$ at 500\,K to $0.81$ at 800\,K. There is significant negative correlation at all temperatures, \new{with $\ln f$ against $1/T$ following a straight line, as shown in the SI, Figure~S1.}

Using just three parameters, the calculated value for $f$ and average values for $\alpha$ and $\nu$, diffusion coefficients (purple pluses) that agree very closely with $D_{msd}$ (blue symbols) can be obtained across a range of temperatures, see Figure~\ref{fig2}. However, the correlated motion dramatically changes the fitted activation energy. A new, and much lower, activation energy ($E_A=0.19\,eV$) is then determined by fitting to the random walk diffusion coefficient, where $\ln D_{rw}=\ln D-\ln f$. This diffusion coefficient would be obtained if individual hopping events were uncorrelated. The activation energy corresponds to the rate-determining step for one diffusive event. \new{If there are multiple barriers along the pathway, the first (lattice-to-interstitial) step will dominate the activation energy. The second continuation (or return) step will have a much lower energy barrier.} \new {This activation energy can be compared to experimental methods that probe single events,} \newer{such as NMR relaxation studies~\autocite{wilkening2012micro}}. \newer{Unlike a NEB calculation, the effects of entropy and collective motion are included, and no prior assumptions are made about the diffusive pathways. Hence, new mechanistic information can be extracted.} \newer{In addition, extracting $f$ in this way also gives a clear picture of how self-correlated motion affects the resulting diffusion coefficient.}


\begin{figure}[h]
\centering
  \includegraphics[height=5.2cm]{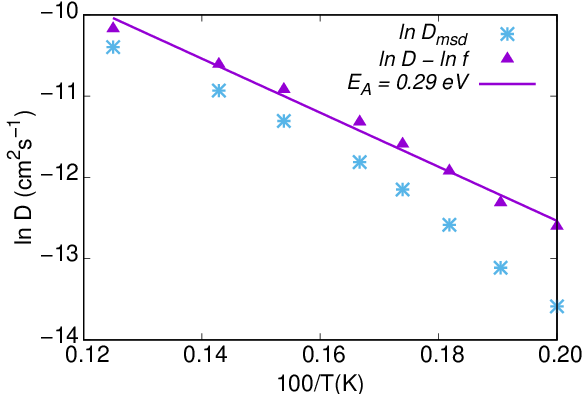}
  \caption{Diffusion coefficients for Li$_6$OCl$_4$. The original diffusion coefficients calculated from the trajectory are shown by blue symbols (gradient of the mean-square displacement). The purple symbols and fitted line show the `random walk' diffusion coefficient with correlation effects removed.}
  \label{fig4}
\end{figure}

\begin{figure}[h]
\centering
  \includegraphics[height=2.6cm]{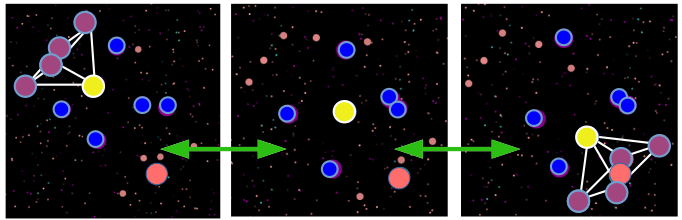}
  \caption{The three images show the movement of a lithium ion (yellow) from one lattice site (left) via an interstitial site (middle) to another lattice site (right). Only lithium (purple), oxygen (pink) and chlorine (blue) neighbouring ions are shown for clarity. This is a rare example of collinear steps.}
  \label{fig5}
\end{figure}

Li$_6$OCl$_4$, a low-dimensional-networked anti-perovskite, does not have a fully ordered structure. The Li$_6$ octahedra are disconnected from each other, \new{see Figure~\ref{fig1} (right), and separated by a `sea' of chlorines.} There is much more flexibility in this structure, but the oxygens continue to provide a reference for lithium movement.  Diffusion coefficients calculated from the mean-square displacement ($D_{msd})$, show a clear deviation from Arrhenius law with temperature, super-Arrhenius behaviour, see Figure~\ref{fig4} (blue symbols).  \new{An equivalent plot for the related Li$_6$OBr$_4$ system is shown in the SI, Figure~S2.} The movement of a lithium ion from one octahedron to another is necessary for long-term diffusivity, the `diffusive step'. Another movement between lattice sites is also possible, corresponding to changes of lattice site within the same octahedron. On their own, these movements cannot contribute to long-term diffusion as there is a maximum distance an ion can travel while remaining within an octahedron. However, such movements may form intermediate movements between the `diffusive steps'. As for Li$_3$OCl, diffusive steps in this system can be identified by considering changes in oxygen neighbours as each lithium in this structure is part of a single octahedron surrounding a central oxygen. Nearly all movements from one octahedron to another take place in two individual steps, moving from a lattice site to an interstitial and then on to another lattice site. However, these steps are not often collinear, a rare example is shown in Figure~\ref{fig5}. During the first step, the lithium ion does not appear to move towards a vacant site, rather away from its site into the sea of chlorines. The subsequent step then depends on a vacant site becoming available nearby, otherwise the ion will return to its initial site. The ease with which the first step is made contributes to the increased diffusion coefficient for this system. At high temperatures, the ion rarely returns to its original site but at lower temperatures, back-and-forth motion is common, with $f$ decreasing from $0.80$ at 800\,K to $0.37$ at 500\,K. The original super-Arrhenius behaviour is caused by the changes in $f$ with temperature, \new{and hence is due to a lack of alternate pathways.} \new{$\ln f$ is a curved line when plotted against $1/T$, as shown in the SI.} When the correlation factor, $f$, is subtracted from the original diffusion constant, $\ln D_{rw}=\ln D_{msd}-\ln f$, an Arrhenius law is recovered from the super-Arrhenius behaviour, see Figure~\ref{fig4}.     

A powerful methodology to analyse MD simulations and extract activation energies for the rate-limiting steps for diffusivity has been demonstrated. Using mechanistic information in the form of ion `hops' and correlations between these hopping events, we are able to extract an underlying Arrhenius form that differs from the original data and \new{demonstrate the effect of different structural geometries and structural flexibility.} For the ordered Li$_3$OCl, this analysis results in an Arrhenius form of a different gradient and an activation energy for a specific mechanism that will enable more accurate comparisons with methods that probe single events \newer{such as NEB and NMR relaxation studies~\autocite{wilkening2012micro}}. For the more flexible Li$_6$OCl$_4$, the original plot has a super-Arrhenius form but extracting correlated motion uncovers an underlying Arrhenius relation for the individual hopping events. \new{The surrounding environment must therefore be responsible for the resulting super-Arrhenius behaviour, evidenced by the increased correlated motion at low temperatures.} This approach can be very generally applied to study ionic transport and has uncovered a wealth of previously unknown mechanistic information for anti-perovskites.

{\it Acknowledgments}\\
The authors thank George Rudman for useful discussions and VKW thanks the Chemistry Department at Durham University for a Career Development Fellowship. This work has made use of the Hamilton HPC Service of Durham University.

{\it Data availability}\\ Our supporting research dataset is published in the Durham University research data repository~\autocite{dataset}.

Supporting Information: correlation factors for Li$_3$OCl and Li$_6$OCl$_4$; diffusion coefficients for the halide-substituted Li$_3$OBr and Li$_6$OBr$_4$; diffusion coefficients for site-to-site transitions in Li$_3$OCl; and additional simulation details.
\printbibliography

\end{document}


\maketitle

\section{Correlation Factors}
\begin{figure}[h]
\centering
\includegraphics[width=0.48\textwidth]{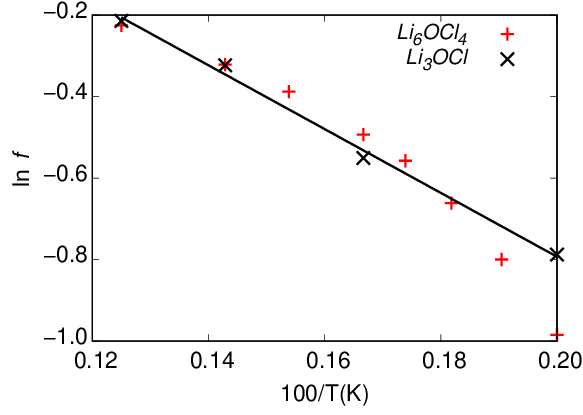}
\caption{Variation of correlation factors ($f$) with temperature for Li$_3$OCl (black crosses, linear) and Li$_6$OCl$_4$ (red pluses, curved).
\label{corrl}}    
\end{figure}

The main text references a correlation factor, $f$, that is calculated for both Li$_3$OCl and Li$_6$OCl$_4$ at different temperatures. Diffusion coefficients for Li$_3$OCl show Arrhenius behaviour (see Figure 2 in the main text) and $\ln f$ is also Arrhenius. For Li$_6$OCl$_4$, the diffusion coefficients show super-Arrhenius behaviour with temperature (see Figure 4 in the main text) and $\ln f$ is super-Arrhenius. The two different correlation factors are shown in Figure~\ref{corrl}.

\section{Halide Substitution}
\begin{figure}[h]
\centering
\includegraphics[width=0.48\textwidth]{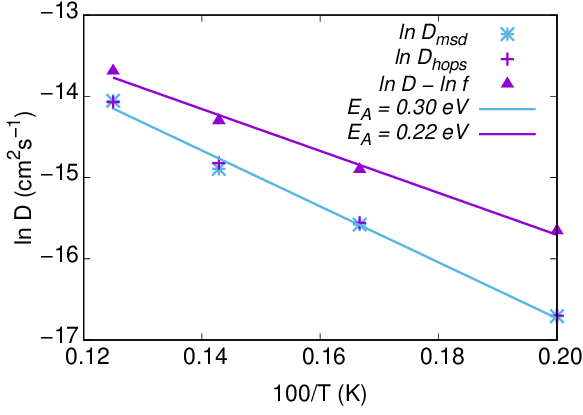}
\includegraphics[width=0.48\textwidth]{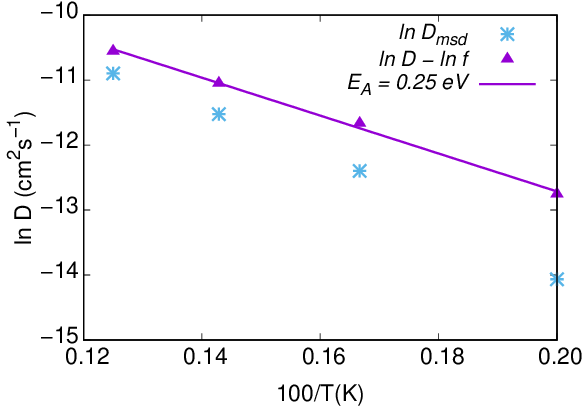}
\caption{Diffusion coefficients for Li$_3$OBr (top) and Li$_6$OBr$_4$ (bottom). The original diffusion coefficients calculated from the trajectory are shown by blue symbols (gradient of the mean-square displacement). The purple triangles and fitted line show the ‘random walk’ diffusion coefficient, (D$_{rw}$), with correlation
effects removed.\label{Brfigs}}    
\end{figure}

To demonstrate that the methodology can be applied more generally, we have repeated the analysis shown in the main text for the similar Li$_3$OBr and Li$_6$OBr$_4$ systems, where bromines replace the chlorines of the original systems. Very similar results for $D_{hops}$ and $\ln D -\ln f$ are obtained as shown in Figure~\ref{Brfigs}.

\section{Site-to-site Transitions}
\begin{figure}[h]
\centering
\includegraphics[width=0.48\textwidth]{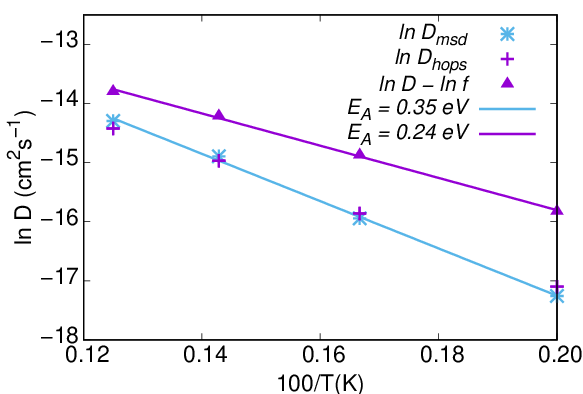}
\caption{Diffusion coefficients for Li$_3$OCl.
 The blue line shows an Arrhenius fit to the original diffusion coefficients calculated from the 10\,ns trajectory mean-square displacement and the purple line shows a fit to the ‘random walk’ diffusion coefficients (D$_{rw}$) with correlation effects removed. In this case due to the increased simulation time and increased observation time (2\,ps) the random walk steps are single-step site-to-site hops.\label{onestep}}    
\end{figure}

There is a choice in how to construct the random walk model for Li$_3$OCl (and Li$_3$OBr). Firstly, as shown in the main text, transitions between sites can be separated into two steps. Alternatively, transitions can be converted into single-step direct transitions between sites and any isolated smaller steps removed. The two random walks are different, with the barrier for the single step being slightly larger as it includes re-crossings of the smaller steps. For the trajectory length (1\,ns) and observation timescale (the time difference between configurations, 0.2\,ps) of results shown in the main text, the two-step process was the natural choice as it retains more information about the diffusive process. We have repeated the analysis over a longer simulation time (10\,ns) and observation timescale (2\,ps). This time, the one-step process dominated and few isolated single steps were observed. The results of the diffusion analysis and the random walk for site-to-site hopping is shown in Figure ~\ref{onestep}.     

\section{Methods}
\subsection{Molecular Dynamics Simulations}
The Molecular Dynamics (MD) simulations were reproduced from those presented in Ref~\cite{dutra2023defect} and performed using LAMMPS~\cite{LAMMPS,Gissinger24}
(\hyperlink{LAMMPS Web Site}{https://www.lammps.org}). It was not our intention to produce new simulation data but we re-ran these simulations in order to save additional configurations along the MD trajectory and obtain locally minimised coordinates. 

These are the details of the simulations run by the authors of Ref.~\cite{dutra2023defect} \newer{and previously developed in Ref~\cite{dawson2018composition}}. A Buckingham potential and additional Coulombic term are used to model the interactions between ions up to a cutoff of 12\,\AA. An additional damping factor is applied to the Coulombic term so that interactions outside the cutoff can be computed using a long-range particle-particle particle-mesh (pppm) solver. Supercells of 10x10x10 unit cells (shown in Ref.~\cite{dutra2023defect}) were constructed and approx. 2\% of the total number of lithium and chloride ions were removed, in equal numbers to ensure charge neutrality. Following minimisation, cell relaxation, and a short (4\,ps) NPT (constant-temperature and constant-pressure ensemble) simulation,  NVT (constant-temperature and constant-volume ensemble) simulations were run with a time step of 2\,fs, with a length of up to 10\,ns.  

In this work, trajectories of 1\,ns were run with ionic configurations saved every 100 time steps (0.2\,ps). \newer{The fitted activation energies for Li$_3$OCl ($0.27$\,eV) and Li$_3$OBr ($0.30$\,eV) are in excellent agreement with those presented in Ref.~\cite{dawson2018composition}, $0.27-0.29$\,eV and $0.29-0.31$\,eV.} In addition, longer trajectories of >10\,ns were run with configurations saved every 1000 time steps (2\,ps). Each of these configurations was minimised to the potential energy surface using the Polak-Ribiere version of the conjugate gradient (CG) algorithm, as implemented in LAMMPS.

\subsection{EL-Cage Analysis}
EL-Cage \hyperlink{EL-Cage code}{https://github.com/VKWard/EL-Cage}~\cite{elcage}, uses a methodology originally described in Ref.~\cite{desouza2008energy}. The code uses a list of nearest neighbours to determine if an ion is mobile. In these systems, the oxygens are used as a reference and movement of the lithium ions can be diagnosed simply by a single change in oxygen neighbour. The first minimum of the radial distribution function is used to create the nearest neighbour list. To detect the loss or gain of a neighbour, a buffer of 1\AA\ is included to ensure that a significant change in the immediate environment of the lithium ion has taken place, such that events involving nearest-neighbours that merely cross the boundary are excluded. The EL-Cage code records all these `cage-breaking' events alongside the number of neighbour changes and the ion labels for the oxygens involved. Post-processing can then be used to produce a random walk with a specific step type from these events. On the timescale of our observations (every 0.2\,ps), site-to-site transitions in Li$_3$OCl can be described by either a two-step or single process. In Li$_6$OCl$_4$, transitions that move a lithium ion from one oxygen neighbour to another generally take place in a two-step process on this timescale. The two-step processes in both systems involve the loss and then subsequent gain of a different oxygen neighbour. A one-step process involves both a loss and then a gain of a different oxygen neighbour. Further details are given in the main text.

\subsection{Diffusion Coefficients}

\begin{figure}[h]
\centering
\includegraphics[width=0.48\textwidth]{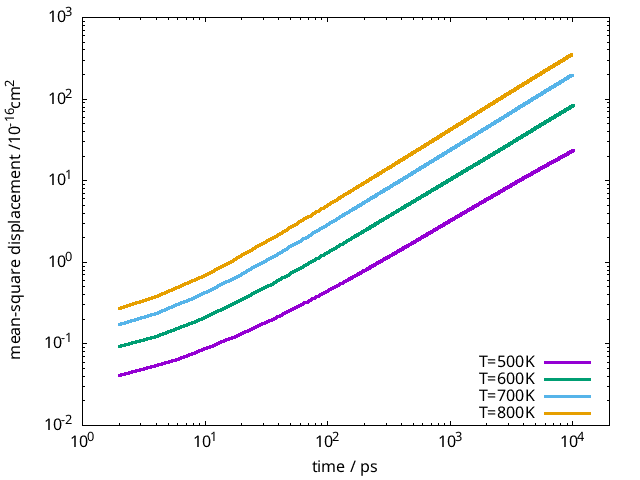}
\caption{Average mean-square displacement for lithium ions in Li$_3$OCl MD simulation trajectories at different temperatures.\label{msd}}    
\end{figure}
\begin{figure}[h]
\centering
\includegraphics[width=0.48\textwidth]{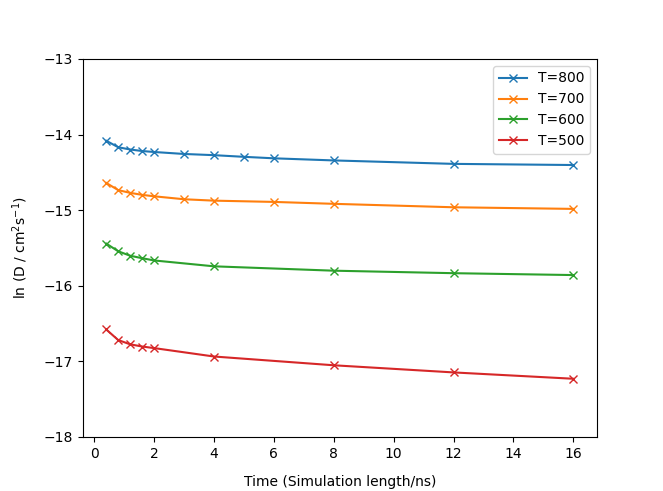}
\caption{Diffusion coefficients for Li$_3$OCl calculated for simulation trajectories of varying length and at different temperatures.\label{Dchanges}}    
\end{figure}

Simulations were examined to ensure equilibration and removing the first 0.5\,ns of each trajectory was found to be more than sufficient. The mean-square displacement (MSD) was calculated for a length of up to half of the remaining simulation trajectory, using an average over all lithium ions and all possible time origins. The MSD was then examined to ensure that a diffusive regime had been reached (see Figure~\ref{msd}) and a long-time gradient was taken to give the diffusion coefficient. \newer{Here, a diffusive regime is identified as the MSD scales approximately linearly with $t$. Within this regime, multiple diffusive steps are taking place and ions are no longer trapped in their local cages.}  

However, in examining the Li$_3$OCl trajectories more closely, it was found that they were not ergodic. In particular, longer simulations produced varying diffusion coefficients, see Figure~\ref{Dchanges}, as the MSD gradient is slowly decreasing. This is not unexpected in a system with significant correlated motion as longer trajectories can reveal longer chains of correlated motion and hence, a lower diffusion coefficient. The diffusion coefficients shown in the main text for Li$_3$OCl should be considered as the diffusion coefficient obtained for a trajectory of that length rather then a long-time value. For Li$_6$OCl$_4$, the diffusion coefficients do converge for $T>500$ and $t>0.5\,ns$. \newer{This does not affect the results of the analysis for $D_{rw}$ as correlated motion is removed nor does it affect the mechanistic information extracted. However, the activation energy obtained for the fit to $D_{msd}$ may increase, and the extracted $f$ values may decrease for longer trajectories. As a diffusive regime has been reached, our analysis of the cage-breaking events that take place within these trajectories is still valid. In fact this analysis provides a robust method for examining non-ergodic trajectories as relevant information can still be extracted.}

\printbibliography